\documentclass[12pt]{iopart}
\usepackage{graphicx}
\usepackage{subfigure}
\usepackage{color,soul}

\newcommand{\Felec}{F_\mathrm{elec}}
\newcommand{\Fdot}{F_\mathrm{dot}}

\newcommand{\Vbias}{V_\mathrm{B}}

\newcommand{\Csub}{C_\mathrm{sub}}
\newcommand{\Ctot}{C_\Sigma}
\newcommand{\Ctip}{C_\mathrm{tip}}

\newcommand{\kB}{k_\mathrm{B}}

\newcommand{\dE}{\Delta E}
\newcommand{\Ec}{E_\mathrm{C}}
\newcommand{\Eadd}{E_\mathrm{add}}

\newcommand{\nshell}{n_\mathrm{shell}}

\newcommand{\ogO}{\omega_0/\Gamma}
\newcommand{\zO}{z_\mathrm{0}}
\newcommand{\Fext}{F_\mathrm{ext}}
\newcommand{\Fexc}{F_\mathrm{exc}}

\newcommand{\dw}{\Delta \omega}

\newcommand{\qdot}{q_\mathrm{dot}}

\newcommand{\Gammap}{\Gamma_{+}}
\newcommand{\Gamman}{\Gamma_{-}}

\begin{document}

% \title[Energy level spectroscopy of quantum dots by force detection]{Energy level spectroscopy of quantum dots by force detection}
\title[Quantum state readout of individual quantum dots by force detection]{Quantum state readout of individual quantum dots by electrostatic force detection}
%\topical{Energy level spectroscopy of quantum dots by force detection}
\author{Yoichi Miyahara, Antoine Roy-Gobeil and Peter Grutter}
\address{Department of Physics, McGill University, 3600 rue University, 
Montreal, H3A 2T8, Quebec, Canada}
\ead{yoichi.miyahara@mcgill.ca}

\begin{abstract}
Electric charge detection by atomic force microscopy (AFM) 
with single-electron resolution
($e$-EFM) is a promising way to investigate
the electronic level structure of individual quantum dots (QD).
The oscillating AFM tip modulates the energy of the QDs,
causing single electrons to tunnel between QDs and an electrode.
The resulting oscillating electrostatic force changes the resonant frequency and
damping of the AFM cantilever, 
enabling electrometry with a single-electron sensitivity.
Quantitative electronic level spectroscopy is possible 
by sweeping the bias voltage.
Charge stability diagram can be obtained
by scanning the AFM tip around the QD.
$e$-EFM technique enables to investigate individual colloidal nanoparticles 
and self-assembled QDs without nanoscale electrodes.
$e$-EFM is a quantum electromechanical system 
where the back-action of a tunneling electron is detected by AFM; 
it can also be considered as a mechanical analog of admittance spectroscopy
with a radio frequency resonator, which is emerging as a promising tool 
for quantum state readout for quantum computing.
In combination with the topography imaging capability of the AFM, 
$e$-EFM is a powerful tool for investigating new nanoscale material systems
which can be used as quantum bits.
\end{abstract}

%\submitto{\NT}
%\maketitle
%\tableofcontents

\section{Introduction}
Quantum dots (QD) are one of the most interesting and
important entities in nanoscience and nanotechnology.
QDs are often called artificial atoms 
as their electronic states become discrete just 
like those in atoms because of quantum confinement in three dimensions.
The emergence of atom-like discrete energy levels 
leads to particular optical and electronic properties of QDs.
Unlike real atoms, one can engineer the size and shape of the QDs,
leading to their tunable optical and electronic properties. 
Therefore, understanding the energy level structure and its
relation to the shape and size of the QD is a key to
understand the properties of QDs, which remains to be elucidated.
For this end, spectroscopic measurement on individual QDs
is essential.
QDs have been attracting much attention in the field of quantum
information processing
as their atom-like discrete energy levels can host charge or spin qubit
and their quantum state can be read out 
by probing the charge state of the QD \cite{Elzerman2004}.
Many studies have already been reported
\cite{Elzerman2005,Petta09302005,Gorman2005}.

Current transport spectroscopy performed in 
single-electron transistor (SET) structures \cite{Fulton1987}
has been instrumental to understanding the electronic levels of individual QDs.
In a SET structure, the energy level of the QD can be shifted
by the gate voltage with respect to the source and drain electrodes. 
When one of the energy levels of the QD
lies in the bias window set by the source-drain bias, 
a single electron can tunnel through two tunnel barriers at the source-QD and the drain-QD.
When the source-drain current is measured 
as a function of gate bias voltage,
peaks appear in the conductance versus gate voltage curves. 
This is known as a Coulomb-blockade oscillation peak.
The energy of these peaks is the sum of 
the energy of the discrete levels
and the Coulomb charging energy of the QD.
A detailed analysis of these spacings thus allows the spectroscopy 
of the electronic energy levels of the QD \cite{Kouwenhoven01}.
Although current transport spectroscopy is very powerful,  
its application has been limited mainly to gate-defined QDs
where a QD is formed in a two dimensional electron gas 
using surface gate electrodes to create a confinement potential.
% in which two-dimensional electron gas typically formed 
% in semiconductor heterostructures
% is further confined laterally by applying bias voltages
% to the surface gate electrodes to create confinement potential to form the QD.

Other interesting classes of QDs such as self-assembled QDs and 
colloidal nanoparticles have not been studied as much
by current transport spectroscopy 
because attaching electrodes to these QDs by lithography techniques
is challenging due to their smaller sizes.
Recently, nanoparticles with more complex shape 
and structure have been developed 
for more functionalities such as biochemical sensing
~\cite{Milliron2004a,Aldaye2007a,Chen2012}.
As attaching electrodes to these nanoparticles and their complexes
is even more challenging, current transport spectroscopy has not been
done in most of these complex structures.

In order to overcome the difficulty, we have developed an alternative 
experimental technique ($e$-EFM).
The charge state of the QD can be detected by electrostatic force microscopy
with sensitivity much better than a single electron charge \cite{Schonenberger1990}.
The AFM tip also acts as a scannable gate, 
thus enabling spectroscopic measurements of single electron charging.
To enable these experiments, the QD needs to be tunnel-coupled 
to a back-electrode.
% The charge state of the QD that are tunnel-coupled to an electrode
% can be controlled by a dc-biased AFM tip positioned near the QD
% which acts as a movable gate.
Oscillating the AFM tip modulates the energy of the QD,
causing a single electron back and forth to tunnel between the QD 
and back-electrode.
The resulting oscillating electrostatic force changes the resonant frequency and
damping of the AFM cantilever, 
enabling the sensitive electrometry with single-electron sensitivity.

As $e$-EFM requires no nanometer scale electrode to be attached to the QD,
it makes it possible to investigate individual colloidal nanoparticles 
and self-assembled QDs which pose major challenges for transport spectroscopy.
This technique is equivalent to admittance spectroscopy with a radio frequency resonator, 
which is emerging as a promising tool for quantum state readout 
\cite{Petersson2010, Petersson2012,Colless2013}.
In combination with the topography imaging capability of the AFM, 
$e$-EFM can be a powerful tool for investigating new nanoscale material systems
which can be used as quantum bits.
The fundamental physics of $e$-EFM is also of great interest 
as the measured interaction between the AFM tip and QD 
can be described as a back-action of a measurements 
on quantum electromechanical system \cite{Clerk2005}.

In this review, we will first describe the basic principle of operation 
of the technique, followed by its application to several experimental systems.
We will then discuss more detailed theoretical aspects which lead to
physical insights which we can gain by this technique.
Finally, we will discuss prospect of the technique, including possible
application to other physical systems.

\section{Single-electron detection by force detection}
\subsection{Overview of  technique}
The experimental technique we describe here is essentially based on electrostatic 
force microscopy (EFM) which is a variant of atomic force microscopy (AFM).
While AFM was shown to be capable of detecting 
a single-electron charge soon after its invention \cite{Schonenberger1990}, 
this single-electron charge sensitivity has not been widely exploited. 

Figure~\ref{fig:setup}(a) depicts the experimental setup of our technique.
 QDs are separated from a conductive substrate (the back-electrode) 
by a tunnel barrier that allows electrons to tunnel 
between the QD and substrate. 
 An oscillating conductive AFM tip is used as a sensitive charge detector 
as well as a movable (scannable) gate. 
 A dc bias voltage is applied between the AFM tip and conductive substrate 
(back-electrode) to control the electron tunneling. 
The tip-QD distance is set to the order of 10~nm 
so that no tunneling is allowed across the vacuum gap between the tip
and QD, making the system a so-called ``single-electron box'' \cite{Wasshuber2001}.
The system with only a single tunnel barrier has several important advantages
both experimentally and theoretically as will be shown later.

%When the dc bias is set to a critical value,
Choice of an appropriate dc bias enables 
a single electron to tunnel back and forth 
in response to the oscillating energy detuning across the barrier 
which is induced by the tip oscillation.
The tunneling single-electron thus produces an oscillating electrostatic force 
which causes peaks both in the resonance frequency 
and the damping (dissipation) of the AFM cantilever.
As is shown later, these peaks are essentially equivalent 
to the Coulomb peaks usually observed in dc transport spectroscopy 
and capacitance/impedance spectroscopy on single-electron transistors
\cite{Kouwenhoven01, Ashoori1993}.

The frequency modulation (FM) mode operation \cite{Albrecht1991} of AFM 
is used to detect the oscillating electrostatic force 
caused by the single-electron tunneling.
Because of the finite tunneling rate, there is a time delay for
the oscillating force with respect to the tip motion.
The in-phase component of the oscillating electrostatic force 
gives rise to the frequency shift signal
and its quadrature component 
to the damping (dissipation) signal.
The effect can also be described as a result of the quantum back-action 
of the tunneling electrons \cite{Clerk2005}.

Another unique feature of this technique is the
scanning capability which enables the spatial mapping 
of single-electron tunneling events.
We will see that the spatial maps of the resonance frequency shift
and dissipation are equivalent to the so-called charge stability diagram 
\cite{Wiel2003, Hanson07}
which provides a wealth of information particularly 
on the systems containing multiple QDs. 
The described technique, termed single-electron electrostatic force microscopy
($e$-EFM), was first demonstrated on QDs formed in carbon nanotube SETs 
\cite{Woodside2002, Zhu05, Zhu08}.
As the present technique requires no patterned electrode to be defined around 
the QDs,
it can open up the spectroscopy on individual QDs of various
kinds that have been very difficult to investigate,
such as colloidal nanoparticle dots and those having complex structures
\cite{Milliron2004a, Chen2012,Swart2016}.

\subsection{Electrostatic force in AFM tip - QD system}
%Figure 1
\begin{figure}
\begin{center}
  \includegraphics[width=150mm]{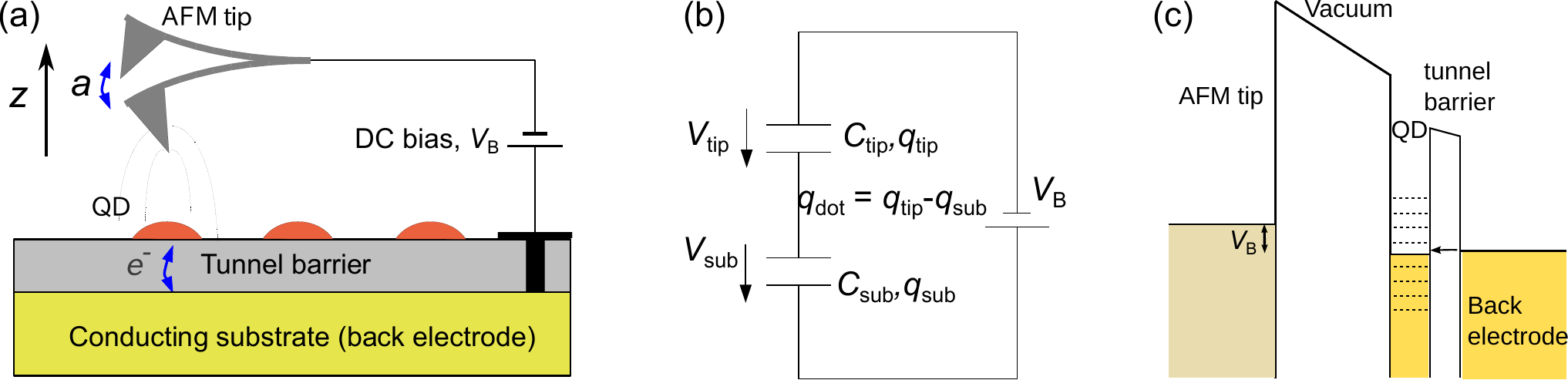}
\end{center}
\caption{(a) Schematic of experimental setup. 
(b) Equivalent circuit of the system shown in (a). 
(c) Energy diagram of the system. \label{fig:setup}}
\end{figure}

Following the equivalent circuit model commonly used 
in the QD transport studies \cite{Wasshuber2001},
we model the system as shown in figure~\ref{fig:setup}(b).
The electrostatic force acting on an AFM tip shown in figure~\ref{fig:setup}(a)
can be calculated by considering 
electrostatic free energy of the system, $W$, 
which is shown below \cite{Stomp05}:

\begin{equation}
\label{eq:Free_energy}
W=\frac{\qdot^{2}}{2\Ctot}
+\frac{\Ctip}{\Ctot}\qdot\Vbias 
-\frac{1}{2}\frac{\Csub \Ctip}{\Ctot}\Vbias ^{2}
\end{equation}
where $\qdot$ is the electrical charge in the QD, $\Ctip$ the tip-QD capacitance, $\Csub$ the QD-back-electrode capacitance and $\Ctot \equiv \Ctip + \Csub$.

Differentiating the electrostatic free energy with respect to the tip position, $z$, 
yields the expression for the electrostatic force as follows:

\begin{equation}
\label{eq:elec_force}
  \Felec = - \frac{\partial W}{\partial z }
= -\frac{1}{\Ctot^2}\frac{\partial \Ctip}{\partial z}\left\{\frac{\qdot^2}{2} 
- \Csub \, \qdot \, \Vbias + \frac{1}{2}\Csub^2 \Vbias ^2\right\}
\end{equation}

The third term on the right hand side 
represents the electrostatic interaction between the charges
in the tip and conducting substrate (back-electrode)
which is known as the capacitive force.
This term is responsible for the parabolic background commonly observed 
in frequency shift versus bias voltage curves.
The second term is proportional to the charge in the QD and 
responsible for the detection of single-electron tunneling.
In the system shown in figure~\ref{fig:setup}(a), 
the charge in the QD, $\qdot=-ne$, ($e$:elementary charge)
is determined by the number of electrons in the QD, $n$,
and can be varied solely by the electron tunneling 
through the tunnel barrier to the substrate
because of the much larger tunnel barrier height of the vacuum gap 
prohibiting the tunneling between tip-QD.

\subsection{Coulomb blockade effect in a single quantum dot system probed by electrostatic force detection}
\label{CB_effect}
In order for a single electron to tunnel into the QD from the back-electrode,
the final state must be more energetically favorable than the initial one.
Considering the electron tunneling process between $n$ electron state and
$n+1$ electron state, 
the free energy of two states needs to be equal to $W(n+1) = W(n)$.
This condition sets the threshold bias voltages for the tunneling to be 

\begin{equation}
\label{eq:threshold_bias}
\Vbias ^{n+1}= \frac{e}{\Ctip}\left(n+\frac{1}{2}\right)
\end{equation}

The separation of two successive peaks, $\Delta \Vbias = e/\Ctip$, 
is proportional to so-called addition energy 
which represents the energy required to add an extra electron to the QD.
In general, the addition energy, $\Eadd$, is determined not only 
by the Coulomb charging energy of the QD, $2\Ec = e^2/\Ctot$, 
but also the energy difference between consecutive quantum states, $\delta$.
$\Eadd$ is related to $\Delta \Vbias$ 
through the relation $\Eadd = e \alpha \Delta \Vbias $ 
where $\alpha$ is the fraction of $\Vbias$ applied across the tunnel barrier
and given as $\alpha = \Ctip/\Ctot$.
$\alpha$ is often called the lever-arm. 
In order to get the true energy scale in the QD, 
$\alpha$ needs to be determined.
As we will see later, $\alpha$ can be determined experimentally from the peak shape 
of the frequency shift or dissipation peak.

As $\alpha$ is determined by $\Ctip$ and $\Csub$,
the oscillating tip leads to the oscillation of the QD energy levels
through the oscillation of $\alpha$.
In other words, the oscillating tip causes the modulation of 
the energy level detuning, $\Delta E$, across the tunnel barrier.
% As we have aleady noted in \ref{CB_effect},  
% only a small fraction of the applied bias voltage,
% $\Vbias$, 
% causes the shift of the QD energy levels 
% because the significant portion of $\Vbias$ drops in the vacuum gap. 
At each $\Vbias^{n+1}$, a single electron can tunnel back and forth 
between the QD and back-electrode
in response to the oscillation in the QD energy. 
It is this oscillating single electron that induces the peaks in frequency shift
and dissipation.
At low temperature, $T$, ($\kB T \ll \Eadd$, $\kB$: Boltzmann constant), 
the number of electron is fixed due to the large addition energy
of the QDs between two adjacent peaks (Coulomb blockade). 

The shape of the peaks can be derived by considering the equation of motion of 
the AFM cantilever that is subject to the oscillating electrostatic force 
caused by the tunneling single electron, $\Fdot$ as described below:

\begin{equation}
  \label{eq:eom}
  m\ddot{z} + m\gamma_{0} \dot{z} + k(z-\zO)=\Fext (z,t) =  \Fdot (t) + \Fexc(t)
\end{equation}

where $m$, $\gamma_0$ and $k$ are the effective mass, intrinsic damping  and spring constant
of the AFM cantilever, respectively. 
$\Fexc$ is the external driving force that is controlled 
by the self-oscillation electronics
\cite{Albrecht1991}.

The second term of Eq.~\ref{eq:elec_force}, $\Fdot$, 
is the back-action of the tunneling process on the detector (\textit{i.e.} AFM cantilever)
and is given by:
\begin{equation}
  \label{eq:Fdot}
  \Fdot(t) = \frac{\Csub}{\Ctot^2}(\qdot \, \Vbias)\frac{\partial \Ctip}{\partial z}
 = -\frac{2\Ec \Vbias}{e}(1-\alpha) \frac{\partial \Ctip}{\partial z}n = -An
\end{equation}
where  $A = -(2\Ec \Vbias/e)(1-\alpha) \partial \Ctip/\partial z$.
$A$ describes the magnitude of the force due to the capacitive tip-QD coupling 
and is dependent on the geometry of the tip and QD and bias voltage.
In practice, the tip-QD distance is a very important parameter
and $A$ increases with decreasing tip-QD distance.

Dynamics of the charge in the QD can be described by the master equation,
${\partial \langle n \rangle}/\partial t = -\Gamma_-(z)\langle n \rangle 
+ \Gamma_+(z)(1 - \langle n \rangle)$
where $\Gamma_+$  ($\Gamma_-$ ) are $z$-dependent tunneling rates 
to add (remove) an electron to the QD 
and $\langle n \rangle$ denotes the average number of electrons on the QD.
The widely accepted treatment of single-electron tunneling implies that 
the operation of single-electron devices that is governed 
by stochastic tunneling events, can be described well 
by time-evolution of the average value \cite{Wasshuber2001}.

\begin{figure}[t]
  \centering
   \includegraphics[height=42mm]{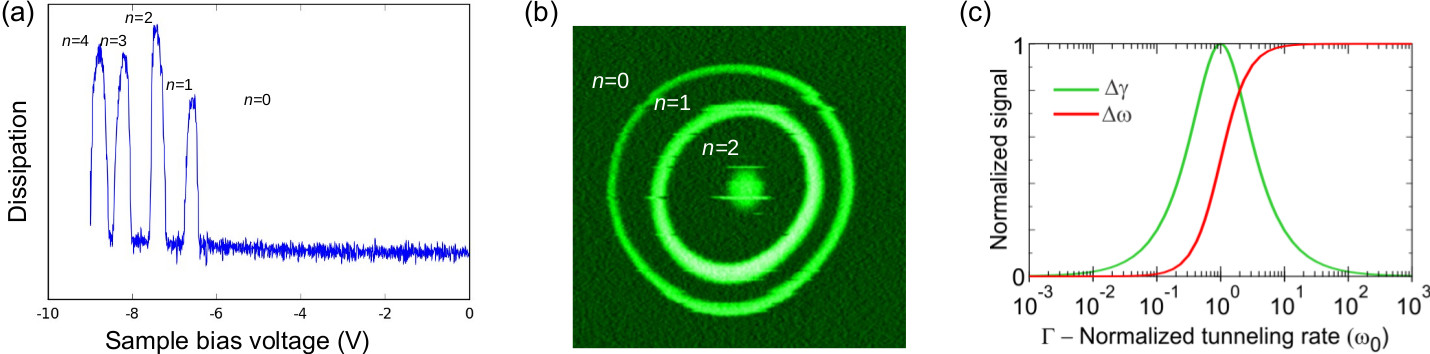}
   \caption{(a)~Single-electron tunneling peaks in dissipation versus bias voltage curve.
(b) Single-electron tunneling rings in the dissipation image taken at a constant bias voltage.
Each peak in (a) corresponds to a ring in (b).
(c) Normalized amplitude of Dissipation and frequency shift 
as a function of tunneling rate. 
\label{fig:diss_peak_and_ring}}
\end{figure}

For a small tip oscillation amplitude,
considering a single non-degenerate level in the QD 
and the linear response of the average charge on the QD, 
the changes in frequency shift, $\dw$,  and dissipation, $\Delta \gamma$,
due to the single-electron tunneling 
are given as follows \cite{Cockins2010a}:

     \begin{equation}
       \dw= -\frac{\omega_0^2 A^2}{2 k \kB T} \frac{1}{1+(\ogO)^2}f(\dE)[1-f(\dE)]
       \label{eq:single_level_deltaf}
     \end{equation}

     \begin{equation}
       \Delta \gamma
       = \frac{\omega_0^2 A^2}{k \kB T}\frac{\ogO}{1+(\ogO)^2}f(\dE)[1-f(\dE)]
       \label{eq:single_level_gamma}
     \end{equation}
where $\omega_0$ is the angular resonance frequency of the cantilever, 
$\Gamma$ the tunneling rate  between the QD and back-electrode 
and $f$ the Fermi-Dirac distribution function.
In this single non-degenerate level case, 
the tunneling rate is independent of the energy 
and equal to tunnel coupling constant.
In general, however, the tunneling rate is energy-dependent and determined 
by the energy-level structure of the QD and back-electrode.
We will see this aspect in the section \ref{section:dos_effect}.
Note that each expression contains a prefactor containing
the ratio between the oscillation frequency of the cantilever
and the tunneling rate.
The tunneling rate can thus be extracted directly by taking these ratio such as 
\begin{equation}
  \label{eq:tunnel_rate}
  \Gamma = -2\omega_0 \frac{\Delta \omega}{\Delta \gamma}
\end{equation}
In other words, the magnitude of $\Delta \omega$ and $\Delta \gamma$ depends 
on the ratio of the cantilever oscillation frequency to the tunneling rate, 
$\omega_0/\Gamma$,
as shown in figure~\ref{fig:diss_peak_and_ring}(c).
The magnitude of $\Delta \gamma$ takes its maximum when $\Gamma \approx \omega_0$.
Note that this determines the condition 
for which both frequency shift and damping signals are observable
\cite{Woodside2002, Zhu08, Brink2007, Tekiel2013}.
Specifically, the tunneling rate between the QD and the back-electrode
needs to be engineered to be similar to the cantilever resonance frequency
(to within a factor of 10).
Note that the resonance frequency of AFM cantilevers 
is typically a few hundred kHz.
As we will see later, this tunneling rate can be engineered
by selecting a barrier of suitable thickness.
In the later sections, we will see that the detailed analysis of the peak shape 
provide such useful information 
as shell-filling of the energy levels in the QD 
and the density-of-states of the QD levels.
%Figure 2
 
% Mention the relation to SGM
%
\subsection{Spatial mapping of charge state}
\label{sec:spatial mapping}
Changing the distance between tip and QD changes the lever-arm 
and thus the energy between QD and back electrode.
Since the AFM tip is a scannable gate and charge sensor,
one can spatially map the charging events,
a unique feature, not available in conventional transport measurements.
The dissipation image shown in figure~\ref{fig:diss_peak_and_ring}(b) demonstrates
the unique capability of this technique.
This image is taken by scanning the tip over a QD 
at a constant tip-height with a constant bias voltage
while acquiring the frequency shift and dissipation signals.
Each of the circular concentric rings in figure~\ref{fig:diss_peak_and_ring}(b)
correspond to a peak in the dissipation versus bias voltage spectrum
obtained at a fixed tip position $(x,y,z)$.

The concentric rings can be understood by considering the energy diagram
of the system.
The energy detuning, $\Delta E$, which governs the electron tunneling
is dependent on the lever-arm, $\alpha$, as well as $\Vbias$.
In fact, $\alpha$ is a function of the tip position such as $\alpha(x,y,z)$
because the tip-sample capacitance depends on the tip position.
If $\alpha$ depends only on the distance between a QD and tip
such as $\alpha(x,y,z)=\alpha(r)$ where $r=\sqrt{(x-x_0)^2+(y-y_0)^2+(z-z_0)^2}$
($(x_0, y_0, z_0)$ denotes the position of the QD) \cite{Tekiel_thesis},
a specific $\Delta E$ allowing tunneling is mapped to a circular ring
around the center of the QD.
Although the real effect of the biased tip on the QD confinement potential
can be more complex, 
this simple model has so far been able to explain 
most of the important features observed in experiments.

\begin{figure}
  \centering
 \includegraphics[width=15.5cm]{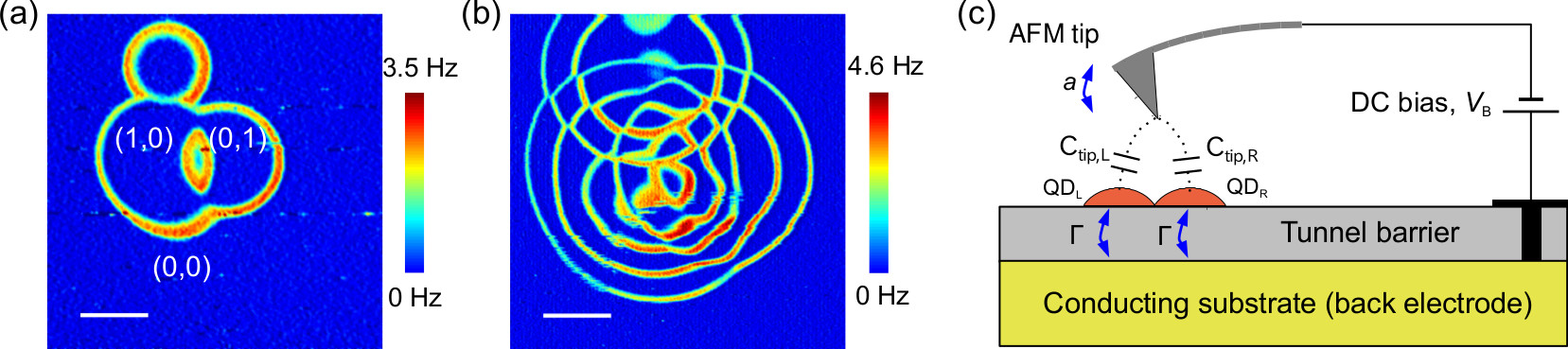}
\caption{(a) and (b) Dissipation images of the same area containing three QDs 
taken at two different bias voltages, $\Vbias = -7.6$~V (a) 
and $\Vbias = -9.0$~V (b).
The numbers in parenthesis in (a)
indicate the number of electrons in the left and right QDs.
Scalebar is 20~nm \cite{Cockins2010a}.
(c) Schematic showing different coupling capacitances to each QD.
\label{fig:multiQD_image}}
\end{figure}

The image indicates that the number of electrons in the QD can be controlled 
by the tip position without changing $\Vbias$.
It is true even for multiple QD complex as shown in figure~\ref{fig:multiQD_image}(a)
because even though there is only one gate (AFM tip),
changing tip position change the lever-arm, $\alpha$, for each QD,
thus the energy of each QD (figure~\ref{fig:multiQD_image}(c)).
Figure~\ref{fig:multiQD_image}(a) and (b) show
the charging of three QDs located close to each other. 
We can see that the number of electrons in each QD 
can be controlled by the tip position.
Furthermore, as expected, we observe
avoided-crossings of the rings,
indicative of the coupling between the QDs.
In fact, these dissipation images are equivalent to the charge stability diagram
\cite{Wiel2003} of multiple QDs as we will see more detail in the later section 
\ref{sec:charge stability diagram}.
Characterizing coupled-QDs is of great interest 
because if the inter-dot tunneling coupling is coherent, 
the QD complex should behave as an artificial molecule
which can host a charge or spin qubit \cite{Petta09302005,Gorman2005}.
%Figure 3

\section{Experimental}
The experiments are performed with a home-built cryogenic 
atomic force microscope \cite{Roseman00}, 
in which a fiber optic interferometer is used for cantilever
deflection sensing \cite{Rugar89, Albrecht1992}.
A fiber-pigtailed laser diode operating at a wavelength of 1550~nm
is used for the interferometer.
The laser diode current is modulated by a radio frequency signal
through a bias-tee to reduce the coherence length of the laser
to reduce phase noise and suppress undesirable interferences.
A typical optical power of $100~\mu$W is emitted from a cleaved 
optical fiber end.
We use commercially available AFM cantilevers (NCLR, Nanosensors)
with typical resonance frequency and spring constant of 160~kHz
and 20~N/m, respectively.
The tip side of the cantilevers are coated 
with 20~nm thick Pt/10~nm thick Ti 
by sputtering to ensure a good electrical conductivity
even at liquid helium temperature.
A dilute helium exchange gas ($\sim 10^{-3}$~mbar) is introduced 
in the vacuum can for the experiments at low temperature
for good thermalization.
The typical quality factor of the cantilevers ranges from 30,000 to 50,000
at 77~K and 100,000 to 200,000 at 4.5~K, 
which is unaffected by the exchange gas.

The fundamental flexural-mode oscillation is controlled either 
by a self-oscillation feedback electronics
which consists of a phase shifter and an amplitude controller
(Nanosurf EasyPLLplus)
or by a digital phase-locked loop oscillation controller (Nanonis OC4).
The resonance frequency shift is measured by a phase-locked loop
frequency detector (Nanosurf EasyPLLplus or Nanonis OC4).
The amplitude of the excitation signal is measured as a dissipation signal.
Care must be taken to correct for the background dissipation signal
caused by the non-flat frequency response of piezoacoustic excitation
systems \cite{Labuda2011}.
 
\section{Epitaxially grown self-assembled InAs QD on InP}
\begin{figure}
\begin{center}
  \includegraphics[height=4.2cm]{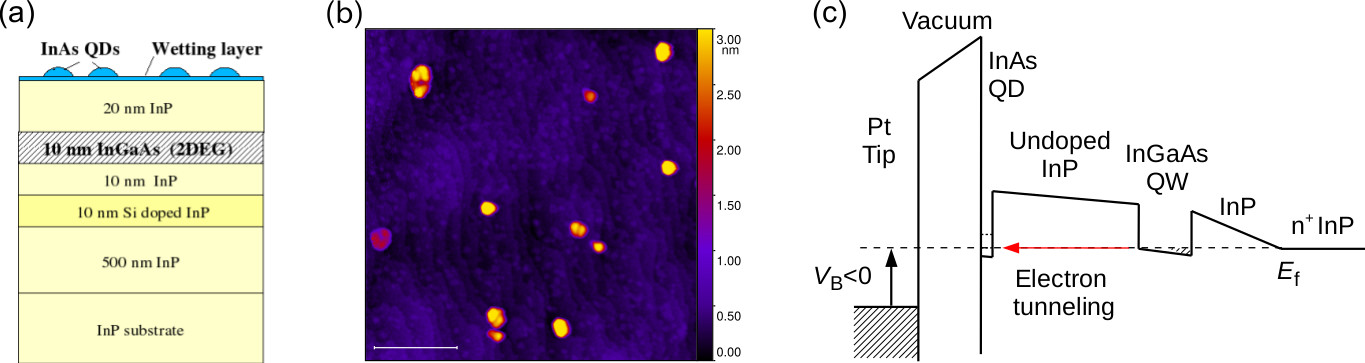}
\end{center}
\caption{(a) Schematic of InAs QD on InP sample structure. 
(b) Tapping mode AFM topogaphy images of the InAs QD sample.
Scale bar is 500~nm.
(c) Energy diagram of the system.}
\label{fig:InAsQD}
\end{figure}

In our first experiment, 
we used epitaxially grown self-assembled InAs QDs \cite{Poole01}.
 The schematic of the structure and the energy level diagram of the sample
are shown in figure~\ref{fig:InAsQD}(a) and (c).
 The two-dimensional electron gas (2DEG) layer formed in a InGaAs quantum well
serves as a back-electrode which works at liquid helium temperature 
and the 20~nm thick undoped InP layer serves as a tunnel barrier.
In this system, InAs islands are spontaneously formed after the growth of monolayer 
thick InAs wetting layer due to the lattice mismatch between InAs and InP.
Figure~\ref{fig:InAsQD}(b) shows an AFM topography image of the sample surface.
The variation in size and shape of the InAs QDs can be clearly seen,
indicating the spectroscopy of individual QDs is highly desired.
%Figure 4

\begin{figure}
\begin{center}
\includegraphics[width=14cm]{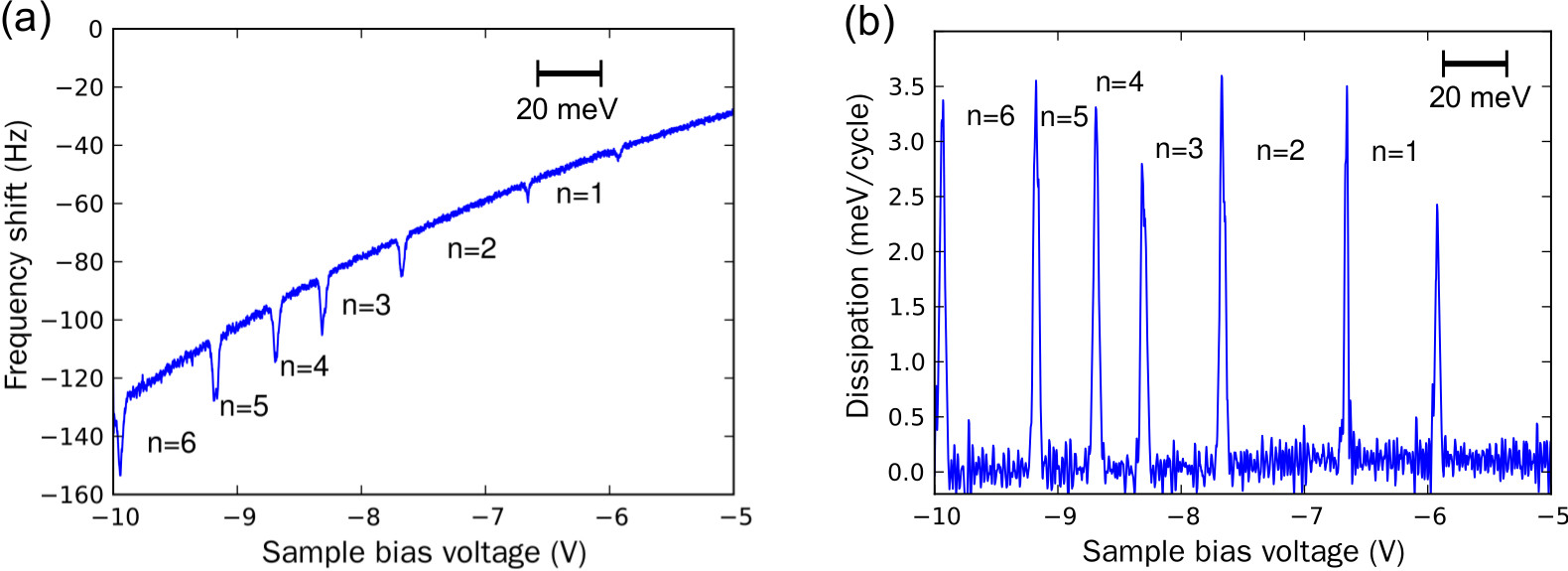}
\end{center}
\caption{(a) Frequency shift and (b) Dissipation versus bias voltage spectra measured 
  on an InAs QD shown in figure~\ref{fig:InAsQD} at 4.5~K.
  The energy scale in the QD  calculated from the lever-arm, $\alpha =0.04$,
is indicated as a scale bar in each figure.}
\label{fig:InAsQD_spectra}
\end{figure}

Figure~\ref{fig:InAsQD_spectra} shows typical bias-spectroscopy curves 
taken over one of the InAs QDs by sweeping the tip-sample bias voltage, $\Vbias$. 
 While sharp dips appear in the frequency shift versus $\Vbias$ curve 
on top of the parabolic background,
sharp peaks appear in the dissipation versus $\Vbias$ curve
at the same bias voltages.
When the sample is negatively biased, the electronic levels 
in the InAs QD are brought down with respect to the back-electrode (InGaAs quantum well)
as shown in figure~\ref{fig:InAsQD}(c).
At a sufficiently high negative sample bias voltage, 
one of these levels is lined up with the back-electrode states
and a single electron can tunnel between the QD and back-electrode.

We can obtain the lever-arm, $\alpha$, experimentally by fitting the observed spectrum
with the theoretical one.
Figure~\ref{fig:Effect_of_degeneracy}(b) and (c) show the results of the fitting.
By assuming the temperature, we can extract the lever-arm, $\alpha=0.04$.
This means only 4~\% of the applied bias voltage 
is applied across the QD-back electrode tunnel barrier
and 96~\% falls off across the tip-QD spacing.
The converted energy in the QD is indicated as a scale bar 
in figure~\ref{fig:InAsQD_spectra}.
Notice that as the temperature broadening becomes smaller at lower temperature,
the oscillation amplitude needs to be reduced accordingly
so that the peak shape is determined not by the tip oscillation
but by the temperature.

We notice the larger peak separation between $n=2$ and $n=3$ peaks and
between $n=6$ and $n=7$. 
It is indicative of the shell structure expected for
2D circularly symmetric potential 
that consists of lowest two-fold degenerate levels 
($s$-shell, 2-spin $\times$ 1-orbital degeneracy)
and the next lowest four-fold degenerate levels ($p$-shell, 2-spin $\times$ 2-orbital degeneracy).
This type of shell structure has previously been observed for an ensemble of 
self-assembled InAs QD \cite{Drexler94, Miller97}.
Assuming this shell structure, we can obtain the charging energy
from this peak separation between peak $n=1$ and $n=2$
as 30~meV using $\alpha=0.04$ obtained before.

However, this simple argument to identify the shell structure is far from convincing.
The more compelling identification usually requires the measurements 
under high magnetic field to acquire the evolution of energy level structure 
(Fock-Darwin spectrum) \cite{Kouwenhoven01, Tarucha1996}. 
In the next section, we will show an alternative way 
for the shell structure identification
based on the asymmetry of the tunneling in and out processes.
%Figure 5

\subsection{Effect of degenerate levels on the dissipation peak}
\label{effect_of_degeneracy}
\begin{figure}
\begin{center}
\includegraphics[width=16cm]{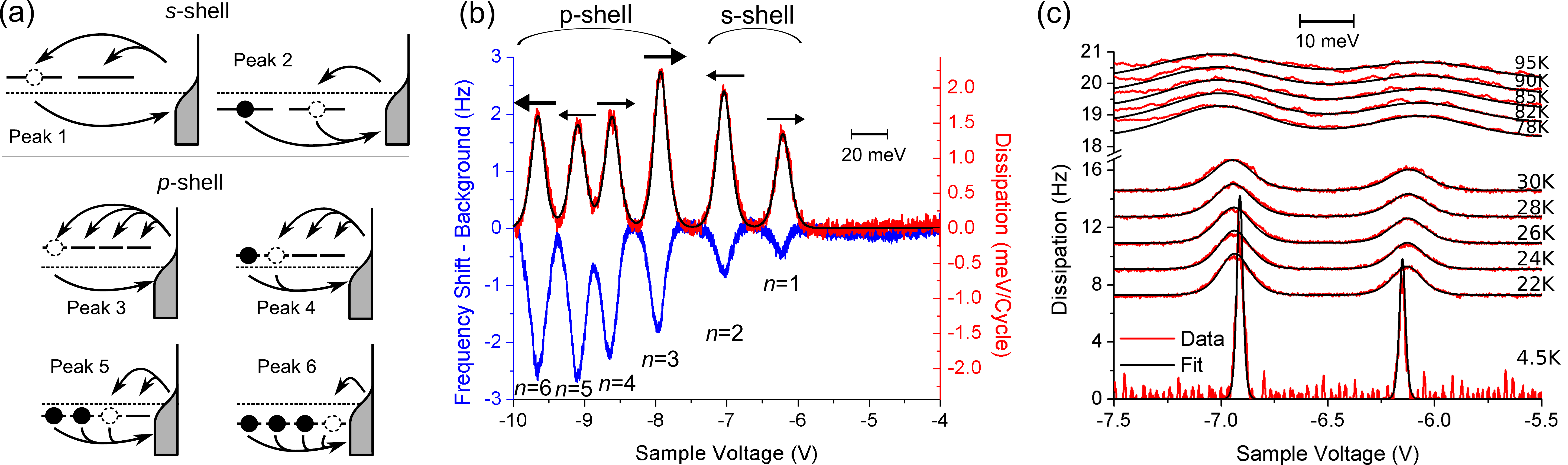}
\caption{
(a) Schematic representation of asymmetric tunneling processes.
Solid horizontal lines depict electronic levels on the QD 
and the grayed area in the right hand side depicts Fermi distribution of 
electrons in the back-electrode. 
The fine dashed line is where the chemical potentials line up
and where a dissipation peak would occur for a nondegenerate level.
(b) The measured dissipation and frequency shift versus bias curves measured at 30~K.
The parabolic background is subtracted from the frequency shift spectrum.
The energy scale in the QD is indicated as a scale bar (20~meV).
The arrows indicate the directions of the peak shifts with increasing temperature.
(c) The measured and fitted dissipation spectra as a function of temperature.
The curves are fitted with Eq.~\ref{eq:diss_small}.
The energy scale in the QD is indicated as a scale bar (10~meV).
Adapted from \cite{Cockins2010a}.\label{fig:Effect_of_degeneracy}}
\end{center}
\end{figure}

When the degeneracy of the energy levels in the QD is taken into account,
the tunneling in and out process is no longer symmetric.
 Figure~\ref{fig:Effect_of_degeneracy}(a) illustrates the asymmetric tunneling process.
When we consider the $n=1$ peak in the spectrum 
shown in figure~\ref{fig:Effect_of_degeneracy}(b),
the peak arises from the transition between $n=0$ and $n=1$ states.
Assuming twofold spin degeneracy under no magnetic field, 
there are two ways for an electron to tunnel into the QD from the back-electrode,
with either a spin-up or a spin-down,
whereas there is only one way for the already occupied electron to tunnel out of the QD.
For the $n=2$ peak, the situation is opposite (one way to tunnel in, two ways to tunnel out).
The same argument for the peaks in $p$-shell 
reveal the similar but more pronounced asymmetry for $n=3$ and $n=6$ peaks 
as illustrated in figure~\ref{fig:Effect_of_degeneracy}(a).

The effect of the asymmetric tunneling process manifests itself in two different ways 
depending on the tip oscillation amplitude.
In the small oscillation amplitude case, 
the positions of the dissipation/frequency shift peaks 
shift with increasing temperature 
in the way that the peaks that belong to the same shell 
repel each other as indicated by arrows
in figure~\ref{fig:Effect_of_degeneracy}(b).
In the large oscillation amplitude case, the peaks that belong to the same shell
get skewed away from each other as shown in figure~\ref{fig:large_oscillation}.
%Figure 6

\subsubsection{Small tip oscillation amplitude case (Weak coupling)}
\label{section:weak_coupling}
In this case, we can obtain the analytical expression 
for the frequency shift and dissipation
by considering the master equation and linear response theory 
\cite{Cockins2010a, Roy-Gobeil2015}.

\begin{equation}
       \dw= -\frac{\omega_0}{2k}\frac{A^2 \Gamma^2}{\kB T} 
       \frac{(\nshell + 1)(\nu - \nshell)}{\omega ^2 +(\phi \Gamma)^2}f(\dE)[1-f(\dE)]
\end{equation}

\begin{equation}
       \Delta \gamma
       = \frac{\omega_0^2 A^2 \Gamma}{k \kB T}
       \frac{(\nshell + 1)(\nu - \nshell)}{\omega ^2 +(\phi \Gamma)^2}f(\dE)[1-f(\dE)]
       \label{eq:diss_small}
\end{equation}
where $\nshell$ is the number of electrons already in the shell before adding a new electron, 
$\nu$ level degeneracy and $\phi = (\nu-\nshell)f + (\nshell + 1)(1-f)$.
While these results indicate the deviation of the peak shape from 
the simple form of $4\cosh^{-2}(\dE/2) = f(\dE)(1-f(\dE))$,
which reflects the asymmetry of the tunneling process,
the deviation is too small to be discernible.
Instead, each peak position changes with temperature in its own unique way
from which the level degeneracy can be identified.

Figure~\ref{fig:Effect_of_degeneracy}(b) and (c) show the direction of the peak shift
of each peak for increasing temperature 
and the temperature dependence of $n=1$ and $n=2$ peaks, respectively.
The peak shift is linearly proportional to the temperature 
and the level degeneracy can be inferred from the slope of the peak shift versus temperature line.
Although a similar effect has been predicated for the current transport measurements 
in single-electron transistors \cite{Beenakker91,Tews2004},  
it has rarely been observed experimentally \cite{Deshpande2000,Bonet2002}. 
The peak shift observed in the dissipation measurement is much more clear
because the effect appearing in the dissipation peak is much more pronounced compared with
the conductance peak as predicted by theory \cite{Cockins2010a}.

\subsubsection{Large tip oscillation amplitude case (Strong coupling)}
\label{section:strong_coupling}
%Figure 7
\begin{figure}
\begin{center}
  \includegraphics[width=15cm]{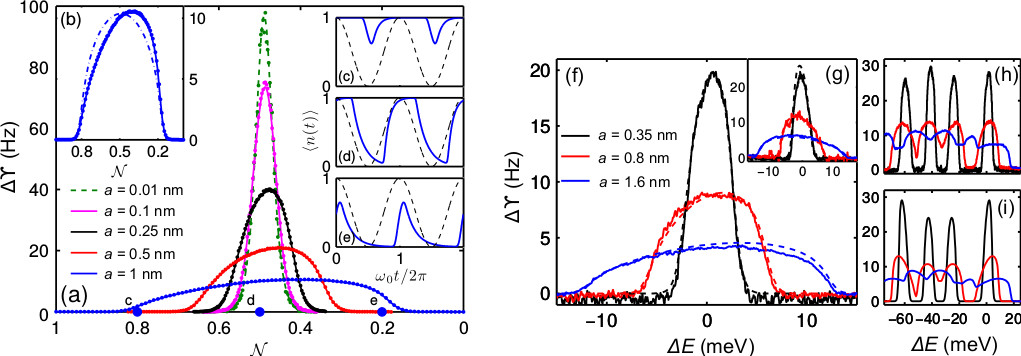}
\caption{(a) Theoretical dissipation peak for $n=1$ 
transition with different oscillation amplitudes. 
$\mathcal{N}=-\Vbias \Ctip/e$ is the dimensionless bias voltage.
Dotted (solid) lines are calculated from simulation (semianalytic theory).
The green dashed line is from linear response.
(b) Adiabatic approximation (dash-dotted), semianalytic theory (solid),
and simulation (dot) for oscillation amplitude $a=1$~nm.
(c)-(e) Average dot charge versus time (solid) for $a=1$~nm, at voltages marked 
in (a). Tip position is shown (thin dashed) as a reference. 
(f) Experimental (solid) and theoretical (dashed) dissipation peaks
for $n=1$ transition with three different oscillation amplitudes.
(g) Dissipation peak for $n=2$ transition.
(h) Experimental dissipation peaks for the transitions in $p$-shell 
with three oscillation amplitudes.
(i) Theoretical dissipation peaks calculated for the peaks in $p$-shell.
Adapted with permission from \cite{Bennett2010}.
Copyright (2010) by the American Physical Society.}
\label{fig:large_oscillation}
\end{center}
\end{figure}

In the large tip oscillation case, the effect of degeneracy appears 
as an asymmetric peak shape
that is much more pronounced and thus clearly observable 
than in the small oscillation amplitude case.
This makes it possible to identify the shell structure by sweeping $\Vbias$ 
with a large tip oscillation amplitude 
than measuring the more challenging temperature dependence 
of the peak position described above.
The large oscillation amplitude refers to the case 
where the maximum change in the energy detuning due to the oscillator, 
$\Delta E_\mathrm{osc} = Aa$ ($a$: oscillation amplitude), 
is comparable to thermal energy, $\kB T$.
In this case, the tunneling rates depend nonlineary on the tip position, $z$
and the master equation needs to be numerically solved to obtain $\langle n(t) \rangle$.
The expressions for frequency shift and dissipation can be obtained
using the theory of FM-AFM \cite{Holscher2001,Kantorovich2004} as shown below:

 \begin{equation}
\Delta \omega = -\frac{\omega_0^2}{2 \pi k a} \int_0^{2 \pi/\omega_0} \Fdot (t) \cos(\omega_0 t)dt
=- \frac{\omega_0^2 A}{2 \pi k a} \int_0^{2 \pi/\omega_0} \langle n(t) \rangle\ \cos(\omega_0 t) dt 
% \Delta \omega = - \frac{\omega_0 A}{2 \pi k a} \int_0^{2 \pi/\omega_0} dt \cos(\omega_0 t)\langle P(t) \rangle\
\label{strong_fshift}
\end{equation}

\begin{equation}
\Delta \gamma = \frac{\omega_0^2 }{\pi k a} \int_0^{2 \pi/\omega_0} \Fdot (t) \sin(\omega_0 t) dt
=\frac{\omega_0^2 A}{\pi k a} \int_0^{2 \pi/\omega_0} \langle n(t) \rangle\ \sin(\omega_0 t) dt
\label{strong_diss}
\end{equation}
Eqs.~\ref{strong_fshift} and \ref{strong_diss} can also be derived 
from a more quantum mechanical approach \cite{Rodrigues2007}.

Figure~\ref{fig:large_oscillation} shows 
the theoretical ((a)-(e)) and experimental ((f)-(g)) 
dissipation versus bias voltage spectra for the $n=1$ transition 
with different tip oscillation amplitudes.
The theoretical curves are obtained from Eq.~\ref{strong_diss} 
using $\langle n \rangle$
calculated from the master equation numerically.
Figure~\ref{fig:large_oscillation}(c)-(e) 
shows the time evolution of $\langle n \rangle$ calculated 
for different bias positions.
We notice that $\langle n (t)\rangle$ is no longer sinusoidal
and that the dissipation peaks become wider and get more skewed 
towards the direction of $n=0$ state for larger amplitude.
The skewed peak stems from the asymmetric tunneling process 
we described above (figure~\ref{fig:Effect_of_degeneracy}(a)).
Figure~\ref{fig:large_oscillation}(f)-(i) show the excellent
agreement between the theory and experiments.
All the theoretical curves are calculated using the following parameters,
$2\Ec=31$~meV, $\alpha=0.04$, and $\Gamma =70, 90$~kHz 
for peaks $n=1$ (figure~\ref{fig:large_oscillation}(f)) 
and $n=2$ (figure~\ref{fig:large_oscillation}(g)),
all of which are obtained from the experimental results taken at weak coupling conditions
(small oscillation amplitude),
only $A$ being a fitting parameter.
The tip oscillation amplitude, $a$, is obtained from the calibrated interferometer signal.
The peaks shown in figure~\ref{fig:large_oscillation}(g) corresponds to the $n=2$ transition 
and is skewed in the opposite direction as $n=1$ peak
because of the opposite asymmetry of the tunneling process.
Figure~\ref{fig:large_oscillation}(h) and (i) show
the experimental and theoretical dissipation peaks versus voltage for the peaks
belonging to the $p$-shell, respectively.
The excellent agreement between the theory and the experiment 
also demonstrates the validity of the theoretical interpretation.
In summary, by finding a pair of peaks which are skewed away from each other,
the shell-filling can be identified 
without measuring magnetic field or temperature dependence of the spectra.

\subsection{Excited-states spectroscopy with a large oscillation amplitude}
\label{section:excited_states}
%Figure 8
\begin{figure}
\begin{center}
\subfigure{\includegraphics[width=14cm]{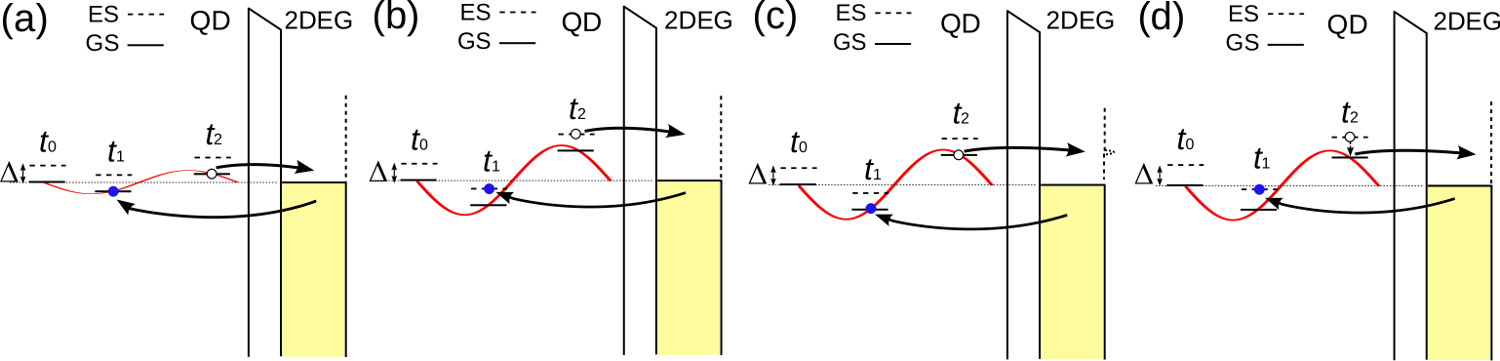}}
\caption{The oscillating AFM tip (symbolized by the red curve), 
modulates the QD levels around. 
The levels are shown at three separate times, 
$t_0$ is the initial position of the levels, at $t_1$ 
and electron tunnels into the dot, and at $t_2$ an electron tunnels out of the dot. 
The filled states of the 2DEG are shown in yellow. 
%The QDand 2DEG are separated by a tunnel junction. 
(a) A small oscillation amplitude opens an energy window that just
allows tunneling into the ground state (GS). 
(b) The larger oscillation amplitude opens an energy window 
that allows tunneling into the first excited state (ES) 
as well as GS ($\Delta$: energy level spacing).
(c) The energy window touches a region where the density of states 
of the 2DEG is not uniform, 
affecting the tunneling rate of the tunneling-out electron. 
(d) Inelastic tunneling process involving photon/phonon emission.
Reprinted with permission from \cite{Cockins2011}.
Copyright (2012) American Chemical Society.}
\label{fig:ESS_principle}
\end{center}
\end{figure}

\begin{figure}
\begin{center}
\includegraphics[width=12cm]{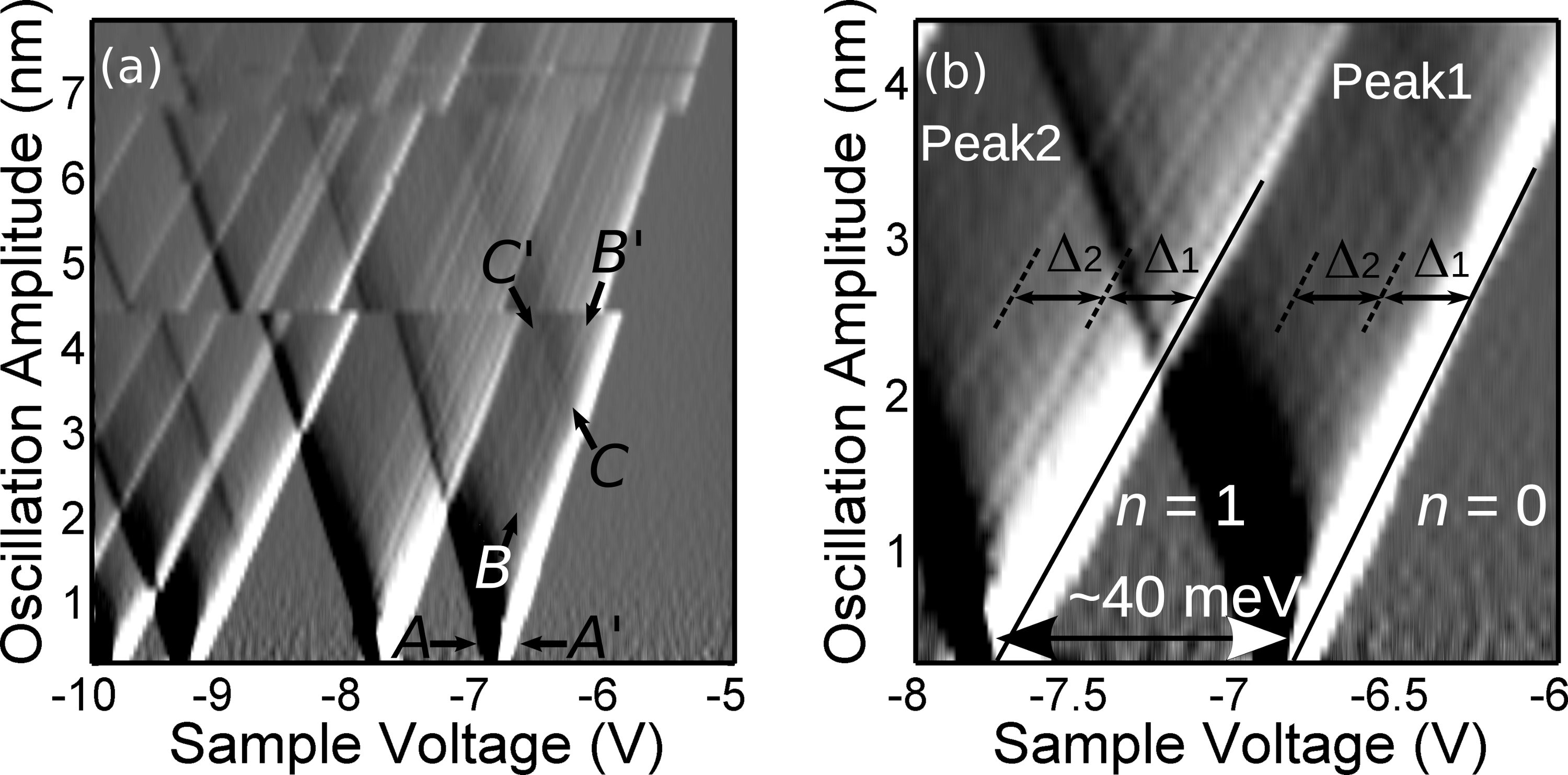}
\caption{(a) Derivative of dissipation versus bias voltage curves taken at different
oscillation amplitudes.
(b) Zoom-up of the region around $A$ and $A'$ indicated in (a).
Reprinted with permission from \cite{Cockins2011}.
Copyright (2012) American Chemical Society.
\label{fig:ESS_results}}
\end{center}
\end{figure}

The analysis in the previous section suggests the possibility of probing the excited states of the QDs
by taking the dissipation spectra with large oscillation amplitudes.
When the tip motion is large enough,
the energy detuning (bias window) modulated by the tip motion 
becomes large enough to allow the tunneling
to the excited states (figure~\ref{fig:ESS_principle}(b)).
This situation is illustrated in figure~\ref{fig:ESS_principle}.
This additional tunneling paths to the excited states show up as additional features
in damping versus voltage spectra.
Figure~\ref{fig:ESS_results} shows such experimental results \cite{Cockins2010a}.
The figure consists of the derivatives of 67 dissipation versus voltage spectra
taken with different tip oscillation amplitudes.

%Figure 9

The down-pointing triangles in the figure are equivalent to the Coulomb diamonds
commonly observed in the current transport spectroscopy experiments 
where the vertical axis is the source-drain bias voltage 
of a single electron transistor (SET) in place of the oscillation amplitude.
The lines that run parallel to the rightmost triangle edges (e.g.,$B$-$B'$)
in figure~\ref{fig:ESS_results}(a) 
result from the opening of additional tunneling paths 
which become available in the tunnel bias window.
In this case, the bias window is set by the lowest and highest energy detuning 
each of which is determined by the closest and furthest position of the tip.
The technique allows us to probe directly the excitation spectrum of the QDs
for a fixed number of electron without optical measurement
and can thus be applied to measure spin-excitation energy under magnetic field 
which is important for single electronic spin manipulations
\cite{Elzerman2005}.
The additional tunneling paths can also be due to inelastic tunneling 
or features in the density of states of the back-electrode 
\cite{Escott2010}.
Ability to probe excited states directly can be applied to such interesting applications
as inelastic tunneling spectroscopy \cite{Natelson2006,Leturcq2009} 
or single-electron spin detection \cite{Elzerman2005}.

\subsection{Spatial mapping of charging on strongly interacting QD
- Charge stability diagram}
\label{sec:charge stability diagram}
\begin{figure}[]
  \centering
  \includegraphics[width=16cm]{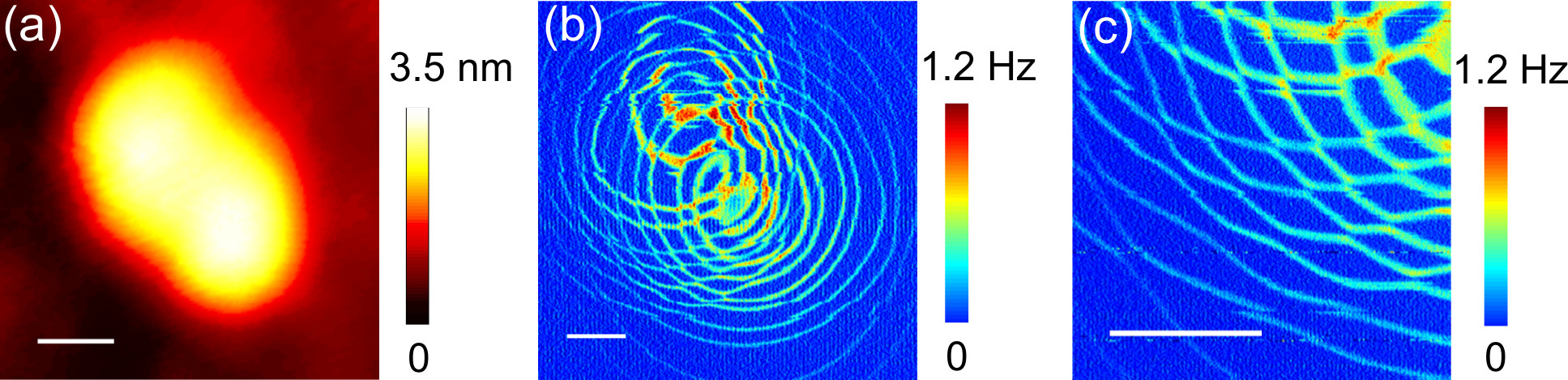}  
  \caption{(a) AFM topography of an InAs QD island. 
    (b) dissipation images of InAs QD taken at $\Vbias = -8.0$~V.
    (c) Zoom up of the lower left corner of the image (b) showing many avoided crossings
  \cite{Cockins2010a}.}
  \label{fig:Strongly_coupled_QD}
\end{figure}

As we have already seen in the section~\ref{sec:spatial mapping}, 
$e$-EFM is capable of obtaining spatial maps of single-electron charging events
in the QD.
Figure~\ref{fig:Strongly_coupled_QD} shows such an example.
Figure~\ref{fig:Strongly_coupled_QD}(a) is the topography image of an InAs island 
and (b) is the dissipation image taken at the same location with $\Vbias=-8.0$~V.
While the topography shows one connected island, 
the dissipation image shows at least two sets of concentric rings,
indicating the existence of multiple QDs in the island. 

Although the current transport spectroscopy on a single QD has been performed on InAs QDs
by attaching a pair of electrodes with a nanogap \cite{Jung05}, 
it would not be straightforward to identify such multiple dots 
just with the conventional conductance versus bias voltage spectra.
The similar spatial maps showing concentric rings have been observed 
by a related technique, scanning gate microscopy (SGM)
\cite{Woodside2002, Pioda04, Bleszynski2007, Boyd2011, Zhou2014}.
In SGM, the AFM tip is used just as a movable gate 
and the conductance of QD devices is measured as a function of the tip position.
While SGM is more widely used to investigate the properties of individual QDs, 
it still requires the fabrication of the wired QD devices.

%Figure 10

Figure~\ref{fig:Strongly_coupled_QD}(c) is the zoom-up of the lower left corner
of figure~\ref{fig:Strongly_coupled_QD}(b).
Overall, the image resembles the charge stability diagram of double QD systems.
The image shows two set of rings which avoid each other.
These avoided crossings are indicative of the coupling between two QDs.
More detailed analysis of the avoided crossing can quantitatively reveal
the nature of the coupling, either capacitive or quantum mechanical (tunnel) 
\cite{Gardner2011},
which is of critical importance for quantum information processing application.

\begin{figure}
  \centering
  \includegraphics[width=16cm]{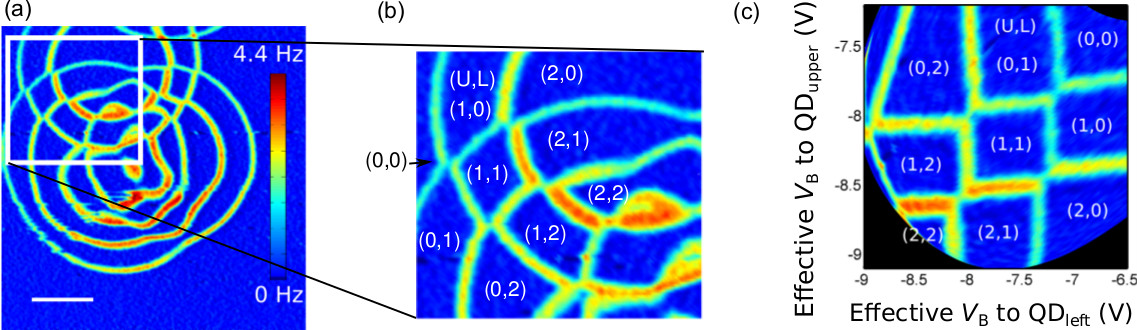}  
  \caption{(a) Dissipation image of the same QD as that in figure~\ref{fig:Strongly_coupled_QD}
    taken with $\Vbias = -8.0$~V at a larger tip-QD distance.
    (b) Zoom-up of (a). 
    (c) Charge stability diagram obtained from (b) by coordinate transformation.
    Vertical axis and horizontal axis are effective bias voltage applied to the upper and left QD,
    respectively. The two numbers in parenthesis in (b) and (c)  indicate the number of electrons 
    in the upper and left QDs.
  \label{fig:converted_charge_diagram}}
\end{figure}

Figure~\ref{fig:converted_charge_diagram} shows 
that a set of concentric rings shown in figure~\ref{fig:converted_charge_diagram}(b) 
can be converted to a conventional charge stability diagram.
Figure~\ref{fig:converted_charge_diagram}(c) can be obtained by the coordinate
transformation, $(x, y) \rightarrow (V_\mathrm{U}, V_\mathrm{L})$
where $V_\mathrm{U}$ and $V_\mathrm{L}$ are equivalent bias 
for the upper and left QDs. 
In this particular case, 
we use the experimentally obtained relations, 
$V_\mathrm{U}=\beta_\mathrm{U} \sqrt{(x-x_\mathrm{U})^2 + (y-y_\mathrm{U})^2 }$
and
$V_\mathrm{L}=\beta_\mathrm{L} \sqrt{(x-x_\mathrm{L})^2 + (y-y_\mathrm{L})^2 }$
where $(x_\mathrm{U}, y_\mathrm{U})$ and  $(x_\mathrm{L}, y_\mathrm{L})$ are
the center of each QD, 
$\beta_\mathrm{U}$ and $\beta_\mathrm{L}$ are constants which can be obtained by
measuring the radius of each ring at different bias voltages, $\Vbias$.

%Figure 11

\subsection{Imaging of Capped QD}
\begin{figure}
  \centering
  \includegraphics[height=4.5cm]{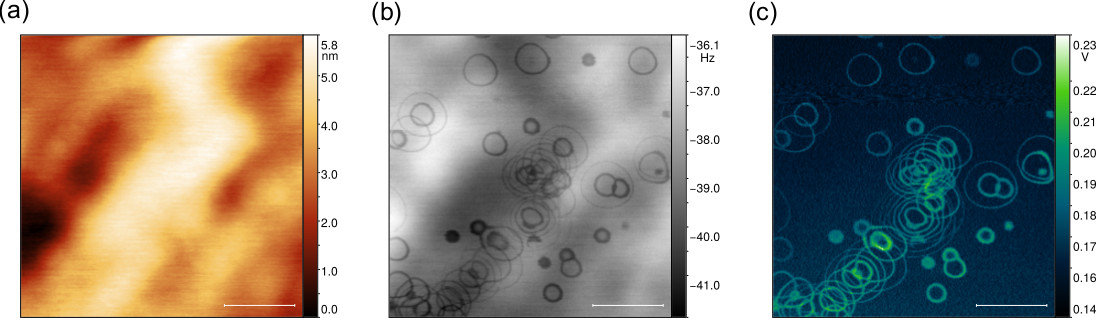}
  \caption{(a) Frequency modulation mode AFM topography, 
    (b) frequency shift and (c) dissipation images of of the
    same location of a capped InAs QD grown on InP taken at $T=4.5$~K. 
    Scale bar is 1~$\mu$m. 
   The topography image was taken separately. The frequency shift and dissipation
   images were taken simultaneously in constant height mode.}
  \label{fig:cappedQD}
\end{figure}

For practical device applications such as quantum dot lasers, 
epitaxially grown semiconductor quantum dots are usually covered 
with a protecting layer \cite{Dalacu09} or even buried in the host matrix
to prevent them from being oxidized.
As $e$-EFM technique relies on the detection of long-range electrostatic force, 
it can be used to probe the capped or buried QDs.
Figure~\ref{fig:cappedQD} shows the topography, frequency shift and dissipation images 
of the InAs QD capped with a thin layer of InP.
The topography image is taken with frequency modulation mode
and the frequency shift and dissipation images are simultaneously 
taken in constant height mode.
While the topography of the QDs is much less obvious than uncapped QDs,
the charging rings appear very clearly in frequency shift and dissipation images.
The background in the frequency shift image originates from the topography 
through the third term of the expression of the electrostatic force in Eq.~\ref{eq:elec_force}.
This demonstrates a clear advantage of the $e$-EFM technique 
over tunneling spectroscopy by STM
which requires a clean surface to ensure a reliable vacuum tunnel gap \cite{Maltezopoulos03}.

%Figure 12

\section{Colloidal Au nanoparticles on alkane-dithiol self-assembled monolayer}
%Figure 13
\begin{figure}
\begin{center}
 \includegraphics[width=14cm]{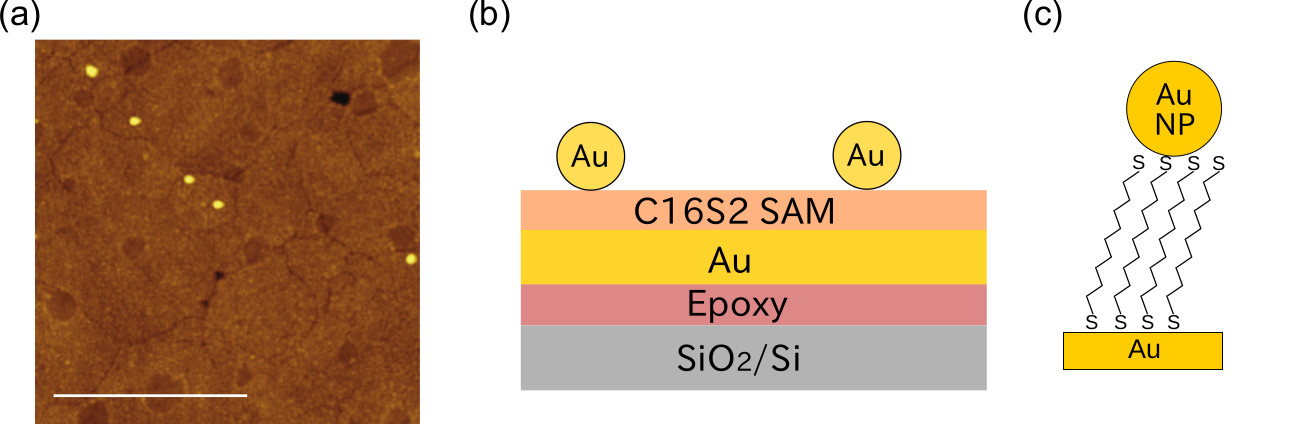}
\caption{(a) AFM topography images of 5~nm diameter Au nanoparticles 
(five bright spots) on C16S2 SAM. Scale bar is 500~nm.
(b) Schematic of the cross sectional view of the sample.
(c) Schematic of Au NP -C16S2 - Au substrate structure.}
\label{fig:AuNP}
\end{center}
\end{figure}

Quantum dots based on colloidal nanoparticles (NP) have been attracting considerable attention
because of their tunable optoelectronic properties by their size and constituent material.
The capability of producing a wide variety of nanoparticles with different materials, 
shapes and sizes together with the possibility of arranging them and 
integrating them with other structures
has a huge potential for various applications \cite{Kim2013}.
Investigating the electronic structure of individual colloidal QDs 
have been of great importance
in order to uncover the relationship between their electronic and geometric structures
and also to engineer more complex structure based on colloidal QDs.
Scanning tunneling microscopy and spectroscopy (STM and STS) have been so far 
used to this end \cite{Swart2016}.
Although the conductance spectroscopy in SET with NP as an island have also been performed
\cite{Kuemmeth08,Nishino2010,Guttman2011}, the device fabrication remains challenging.
%Figure 14
\begin{figure}
  \includegraphics[width=15cm]{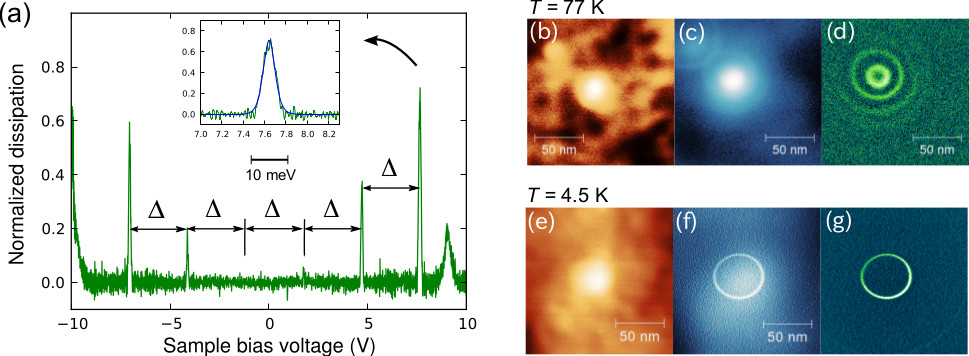}
\caption{(a) Dissipation Spectrum on Au NP taken at $T=4.5$~K. 
The energy scale in the QD is indicated as a scale bar (10~meV).
(Inset) Blue solid curve is a fitted curve with the theory.
(b) and (e) Topography, (c) and (f) frequency shift, (d) and (g) dissipation 
images on Au NP taken at 77~K and 4.5~K, respectively.
\label{fig:AuNP_results}}
\end{figure}

We apply $e$-EFM technique to study the single-electron charging 
in colloidal bare gold nanoparticles (Au NP).
For this study, the sample as shown in figure~\ref{fig:AuNP}(b) has been prepared.
The self-assembled monolayer (SAM) of 1,16-hexadecanedithiol (C16S2) is formed 
on a template strip gold surface and is used as a tunnel barrier.
Bare Au NPs with 5~nm diameter are deposited on the C16S2 SAM layer to form
a Au NP-C16S2-Au junction (figure~\ref{fig:AuNP}(c)).
Figure~\ref{fig:AuNP}(a) shows the AFM topography image of the prepared sample.
Covalent Au-S bonds form stable links between Au NPs and Au substrate through
alkyl chains that act as a tunnel barrier \cite{Akkerman2007,Song2010}.
The alkanedithiol SAM is ideally suited to optimize the tunnel barrier so that
the tunneling rate matches the oscillation frequency of the AFM cantilever.

Figure~\ref{fig:AuNP_results}(a) shows a typical dissipation versus bias voltage curve
taken above a 5~nm Au NP on the sample at the temperature of 4.5~K.
A set of very sharp single-electron charging peaks are clearly seen in the figure.
The equal separations between the neighboring peaks indicate that the Au NP is 
in the classical Coulomb blockade regime ($\delta \ll \kB T \ll \Ec$),
which is consistent with the expected $\delta \approx 1$~meV estimated for 5~nm Au NPs
\cite{VonDelft2001}.
The inset shows the zoom up of the rightmost peak together with the fitted curve 
(blue line) with the function, $\cosh^{-2}[e\alpha(\Vbias-V_0)/2\kB T]$
from which $\alpha=0.0054$ is extracted. 
Using the extracted $\alpha$ value, the charging energy of the Au NP is obtained
as $\Ec = 16$~meV.

Figure~\ref{fig:AuNP_results}(b-g) shows the spatial mapping results on a Au NP.
The top and bottom panel shows the result at 77~K and 4.5~K, respectively.
The concentric rings due to the single-electron charging can be clearly seen 
even at 77~K but the contrast of the ring to the background is much higher at 4.5~K.
As is shown here, $e$-EFM technique can be used to investigate the nanoparticle
complex as those demonstrated in Ref.~\cite{Milliron2004a,Aldaye2007a,Chen2012}.

\subsection{Revealing density-of-states from tunneling-rate spectrum}
\label{section:dos_effect}
The expressions for $\Delta \omega$ and $\Delta \gamma$  
given in Eq.~\ref{eq:single_level_deltaf} and \ref{eq:single_level_gamma} 
are derived for a single non-degenerate level,
in which case the total tunneling rate, $ \Gamma_{\Sigma} = \Gammap+\Gamman$, is constant 
and equal to the tunnel coupling constant.
In general, the tunneling rate depends on the energy level structure 
and is thus energy dependent.
The corresponding results for generic energy-dependent tunneling rate 
in the limit of weak coupling (small oscillation amplitude) 
are obtained as follows \cite{Roy-Gobeil2015}:

\begin{equation}
\label{fshift}
\Delta \omega = -\frac{\omega_0 A^2}{2 k} \frac{(\Gammap' \Gamma_{\Sigma} - \Gammap \Gamma_{\Sigma}')}{(\Gamma_{\Sigma}^2 + \omega^2)}
\end{equation}

\begin{equation}
\label{diss}
\Delta \gamma = \frac{\omega_0^2 A^2}{k \Gamma_{\Sigma}} \frac{(\Gammap' \Gamma_{\Sigma} - \Gammap \Gamma_{\Sigma}')}{(\Gamma_{\Sigma}^2 + \omega^2)}
\end{equation}
where $'$ denotes derivative with respect to energy. 
The total tunneling rate between the QD and the back-electrode, $\Gamma_{\Sigma}$,
can be directly measured as a function of the electrochemical potential detuning 
by noting that:

\begin{equation}
\label{total_trate}
\Gamma_{\Sigma}(\Delta E) = -2 \omega_0 \frac{\Delta \omega}{\Delta \gamma}
\end{equation}

This points to the interesting possibility of tunneling-rate spectroscopy
from which the energy level structure (density-of-states) of the QD can be extracted. 
It should be noted that the measuring the tunneling rate with current spectroscopy measurement
in SET is not as straightforward because two tunnel barriers are involved.
Alternative single-electron counting techniques require fast charge sensors \cite{Muller2012}.

\begin{figure}
\begin{center}
\includegraphics[width=15.5cm]{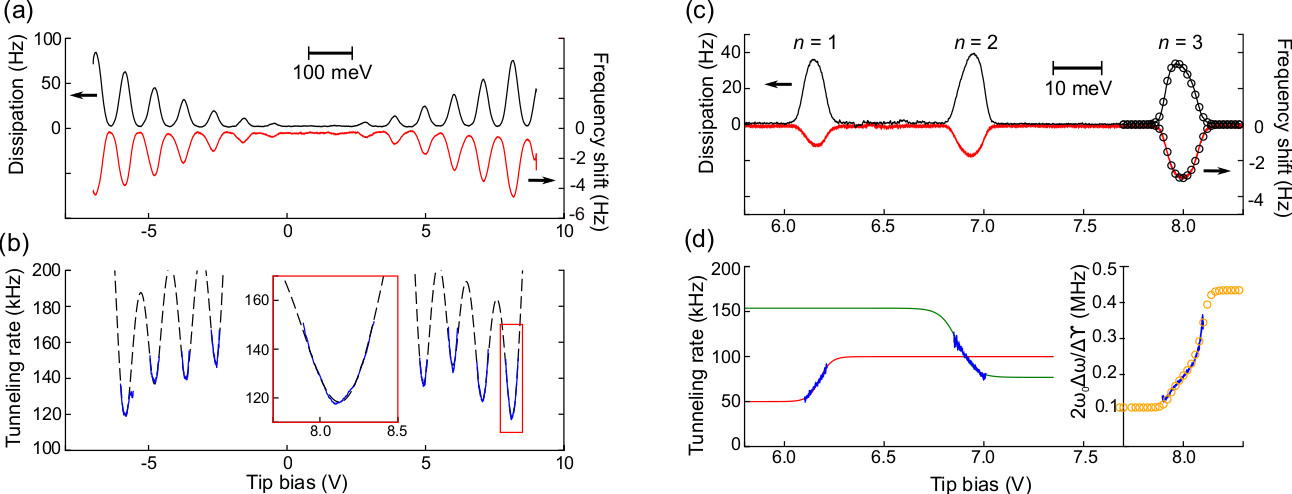}
\caption{Effect of density-of-states on tunneling rate. 
(a) Experimental frequency shift and dissipation versus bias voltage curves
measured on Au NP at 77~K.
The energy scale in the Au NP is indicated as a scale bar (100~meV) 
determined from the lever-arm, $alpha=0.064$.
(b) Extracted tunneling rate from the data above using Eq.~\ref{total_trate}
(blue solid line).
Dashed line is a fit to the analytical expression for a continuous density of states.
(c) Experimental frequency shift and dissipation versus bias voltage curves
measured on InAs QD at 4~K.
The energy scale in the QD is indicated as a scale bar (10~meV) 
determined from the lever-arm, $alpha=0.04$.
(d) Corresponding tunneling rate data (blue) obtained from the data above.
Green and blue solid lines are fits to the tunneling rate expected 
for 2-fold degenerate levels.
Circles represent a best fit solution assuming tunneling 
into an empty 4-fold degenerate level accounting for the strong coupling effect.
Adapted with permission from \cite{Roy-Gobeil2015}.
Copyright (2015) American Chemical Society.}
\label{fig:DOS_effect}
\end{center}
\end{figure}

Figure~\ref{fig:DOS_effect} shows three examples of such tunneling-rate spectroscopy using $e$-EFM.
Figure~\ref{fig:DOS_effect}(b) shows energy dependent tunneling rate for a Au NP 
which is obtained from the $\Delta \omega$ and $\Delta \gamma$ spectra 
shown in figure~\ref{fig:DOS_effect}(a).
A parabola like energy dependent tunneling rate (blue curve) is clearly seen in the spectrum 
which is fitted with the energy dependent tunneling rate expected 
for the continuous density of states (dashed line) \cite{Brink2007, Beenakker91}. 
Figure~\ref{fig:DOS_effect}(c) shows the experimental frequency shift and
dissipation spectra measured on InAs QD at 4.5~K.
(similar to the result in figure~\ref{fig:InAsQD_spectra} but the bias axis is reversed).
Three peaks for $n=1$, $n=2$ and $n=3$ are seen from left to right.
Figure~\ref{fig:DOS_effect}(d) shows the corresponding tunneling rate spectra for each peak
(blue line).
The green and red lines are fits with the the theoretical tunneling rate expression expected
for 2-fold degenerate levels with the shell-filling, $\nshell=0$ and $\nshell=1$, respectively.
The yellow circle is a fit to the theory for 4-fold degenerate levels with $\nshell=0$.
We can clearly see the asymmetric tunneling processes which is discussed 
in section~\ref{effect_of_degeneracy} in the tunneling rate spectra
although it is not obvious at all to identify them 
in the charging peaks in either frequency shift or dissipation spectrum.
In other words, it is possible to identify the shell filling 
from the slope of the tunneling rate spectra.
This approach can be extended to identify more general energy level structures
 \cite{Roy-Gobeil2015}.

%Figure 15

% As we have seen in the previous sections \ref{}, 
% the energy level structure of the QDs can appear in the dissipation/frequency shift spectra
% in several different ways.

\section{Au island on a few monolayer thick NaCl grown on
Fe(001) surface}
The charge transfer process to metallic islands on insulator surfaces have been 
the subject of active research particularly in the context of supported model
catalysts \cite{Henry2015}.
STM has been a main tool for investigating the relationship between the size and
shape of individual islands and their electronic properties including charge transfer
\cite{Giordano2011}.
However, as STM requires a dc current typically higher than 1~pA,
it can only be applied to systems of a few monolayers of insulators.
Here, we demonstrate that $e$-EFM technique can be applied to 
the gold islands deposited onto thicker insulating films.

\begin{figure}
  \centering
  \includegraphics[width=15cm]{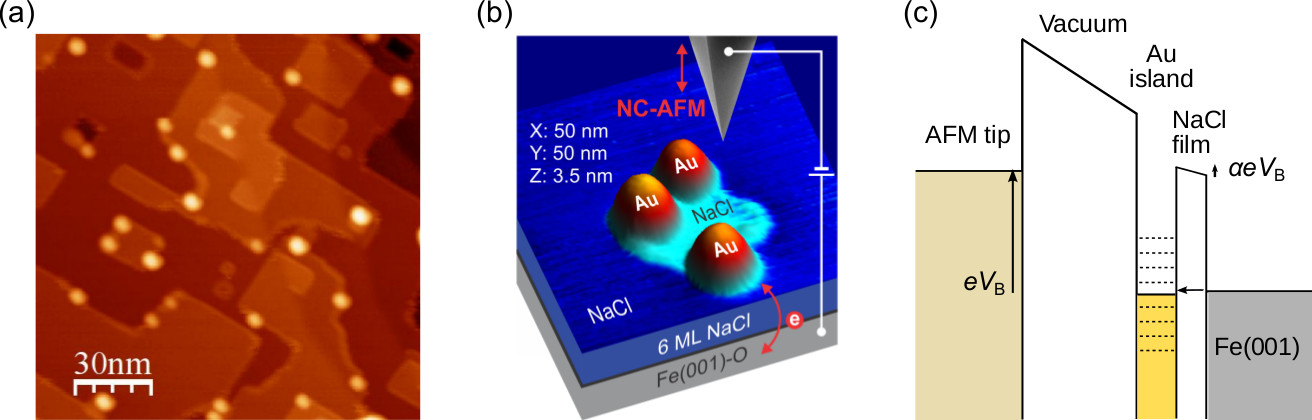}
  \caption{(a) Frequency modulation mode AFM topography of Au islands on 7 ML thick NaCl.
    (b) Schematic diagram of the experimental setup.
    (c) Energy level diagram of the system. 
Adapted with permission from \cite{Tekiel2013}.
Copyright (2012) American Chemical Society.}

  \label{fig:Au_island_schematic}
\end{figure}

Figure~\ref{fig:Au_island_schematic}(a) shows the AFM topography image of the sample
with a 7~ML thick NaCl grown on a Fe(001) surface with a submonolayer coverage of Au.
The NaCl layer grows in a nearly perfect layer-by-layer growth mode \cite{Tekiel2012},
allowing for the tunneling rate to be controlled.
Figure~\ref{fig:Au_island_schematic}(b) and (c) show the schematic and the energy level diagram  
of the experimental setup.
The size of the studied Au nanoislands is 3.5~nm in height and 10~nm in apparent diameter
which is likely to be larger than the actual dimension of the QDs
due to tip convolution effects.

\begin{figure}[t]
  \centering
  \includegraphics[width=15.5cm]{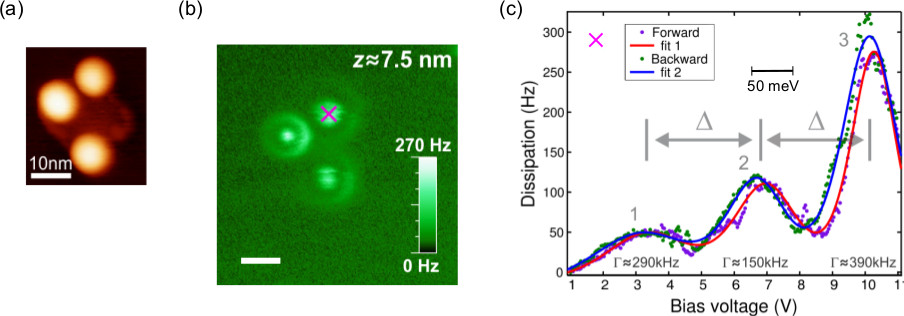}
  \caption{(a) Frequency modulation mode AFM topography image of Au islands on 7 ML thick NaCl.
    The height of the islands is 3.5~nm.
    (b) Dissipation image on the same Au islands taken at $\Vbias=10$~V and $a=1$~nm.
    (c) Dissipation versus $\Vbias$ curve taken on the Au island indicated as $\times$ in (a).
The energy scale in the Au NP is indicated as a scale bar (50~meV) 
determined from the lever-arm, $\alpha=0.04$.
    Adapted with permission from \cite{Tekiel2013}.
    Copyright (2012) American Chemical Society.}
  \label{fig:Au_island_results}
\end{figure}

Figure~\ref{fig:Au_island_results}(b) shows the dissipation image taken over the three Au
islands shown in figure~\ref{fig:Au_island_results}(a) with the tip height of 7.5~nm
and $\Vbias=10$~V. 
The measurements were performed at room temperature under ultra-high vacuum condition.
A circular ring with a dot at its center appear for each of the islands,
indicating single-electron tunneling between the Au islands and Fe(100) back-electrode
through 7~ML NaCl film \cite{Tekiel2013}.
Figure~\ref{fig:Au_island_results}(c) is the dissipation versus bias voltage spectrum 
taken at the center of one of the three islands (indicated by $\times$).
We can see three single-electron charging peaks whose separations are roughly equal.
Fitting the peaks with $\cosh^{-2}[e\alpha(\Vbias-V_0)/2\kB T]$ provides $\alpha=0.04$
by which the addition energy of $137\pm27$~meV can be obtained.
By taking this value as the charging energy of the Au island
(assuming the negligible contribution of the quantum confinement energy),
we can determine the shape of the Au island as a truncated sphere with the base diameter
of 4.2~nm from the finite-element electrostatic modeling 
of the system using the measured height of the island.
Similar charging peaks are observed in the frequency shift versus bias voltage spectrum
as well, enabling the determination of the electron tunneling rate through NaCl layer ranging 
from 61~kHz to 285~kHz for the three peaks.

The approach presented in this section can be applied to a variety of interesting systems
such as metallic islands on oxide films that are widely studied model catalysts.
The interaction of even smaller islands with molecules could be investigated 
by monitoring the electronic level structure of these islands with the technique discussed above.

%Figure 16

%Figure 17

\section{Possible application to other systems}
The application of $e$-EFM technique is not limited to QDs. 
Single-electron transistors incorporating individual molecules have been studied actively 
in recent years \cite{Natelson2006},
and different charge states of single molecules have been observed 
in conductance spectroscopy \cite{Park2000,Park2002,Kubatkin2003}.
We can therefore imagine that $e$-EFM can be applied to investigate the properties of 
single molecules.
As the fabrication of single-molecule SETs is inherently challenging \cite{Yu2004}
because a single molecule cannot be placed in the nanogap electrodes 
in a well controlled manner,
$e$-EFM could be used as a new technique to investigate 
the electronic energy levels of individual molecules 
that interact with a substrate \cite{Kubatkin2003}.

Another interesting system for $e$-EFM technique is individual dopant atoms in semiconductors.
While the behavior of individual dopant atoms is becoming increasingly important in ever shrinking 
semiconductor electronic devices,
these individual dopants are emerging as a new class of quantum mechanical entities
whose discrete energy levels can be used for applications in quantum information processing 
such as qubits and non-classical light sources \cite{Koenraad2011}.
Although spectroscopy of individual dopants can be performed in a single-electron transistor
geometry \cite{Sellier2006, Fuechsle2012}, 
STM has also been applied to detect the charge state of individual dopant 
by detecting the electron tunneling through bound states due to the dopant atoms
\cite{Teichmann2008a}.
In addition to the advantage of STM based experiments 
which do not require the fabrication of device, 
$e$-EFM can be applied to the dopants that are much more weakly coupled 
to the bulk states 
as the required tunneling rate to observe a signal is much lower than that for STM
(recall that 1~nA corresponds to $10^{10}$ electrons per second).

\section{Relation to other techniques}
The $e$-EFM technique shares several common features with capacitance/admittance 
spectroscopy.
Single-electron tunneling spectroscopy has been performed 
with capacitance spectroscopy technique \cite{Drexler94, Ashoori92}.
In capacitance spectroscopy, a QD is connected to a lead through a tunnel barrier 
and the charge in the QD is controlled by a gate voltage. 
Single-electron tunneling is induced by applying an ac voltage to the gate 
and the resulting ac current is measured by a lock-in amplifier.
Although the capacitance spectroscopy has successfully been applied 
to gate-defined QDs \cite{Ashoori1993, Ashoori92, Ashoori93}, 
it has not been widely adapted for individual QDs 
because of the difficulty in detecting 
the weak electrical signal from single-electron tunneling.
Recently, similar measurements using radio frequency (RF) resonators 
have emerged as an alternative technique to current transport spectroscopy
for probing the quantum states of QDs.
In this scheme, an RF resonator is directly coupled to the QD 
through either a source/drain electrode 
\cite{Petersson2010,Villis2011a,Chorley2012, Frey2012, Hile2015}
or a gate electrode 
\cite{Colless2013, Verduijn2014,GonzalezZalba2015, 
Gonzalez-Zalba2016, Frake2015}
and the changes in the phase and amplitude of the reflected RF signal 
due to the single-electron tunneling are detected (RF reflectometry). 
The changes in the phase and dissipated power of the microwave signal are expressed 
for small RF excitation limit as follows: 
\begin{equation}
  \label{eq:phase_response}
  \Delta \phi \simeq -\frac{\pi Q_\mathrm{res}}{C_\mathrm{p}} \frac{(e\alpha)^2}{2\kB T}
  \frac{1}{1+(\omega_0/\Gamma)^2} f(\dE)[1-f(\dE)]
\end{equation}

\begin{equation}
  \label{eq:capacitance}
  \Delta P \simeq  \frac{(e\alpha V_\mathrm{g}^\mathrm{rf})^2}{2\kB T}
  \frac{\ogO}{1+(\ogO)^2}f(\dE)[1-f(\dE)]
\end{equation}
where $Q_\mathrm{res}$ is the quality factor of the RF resonator, 
$C_\mathrm{p}$ the paracitic capacitance,
$\alpha$ is the lever-arm, $\alpha = C_\mathrm{g}/C_\Sigma$, 
$V_\mathrm{g}^\mathrm{rf}$ the RF signal excitation amplitude \cite{GonzalezZalba2015}.

This technique is emerging as a promising way to detect the charge state of QDs
which is indispensable for readout of the qubits based on QDs.
$e$-EFM can be considered as a mechanical analog of this technique.
In $e$-EFM, a mechanical resonator (AFM cantilever) 
is capacitively coupled to the QDs in place of an RF resonator 
and its response to the change in charge state is detected as 
the changes in resonance frequency and dissipation, 
similarly to the RF admittance measurements.
Therefore, admittance spectroscopy should be able to measure
shell-filling (\ref{section:weak_coupling}, \ref{section:strong_coupling}), 
perform excited-states spectroscopy  (\ref{section:excited_states}) and
determine density-of-states (\ref{section:dos_effect}) 
in complete analogy to our $e$-EFM measurements.
Although RF reflectometry techniques have been applied for other systems 
than gate-defined QDs
such as single dopants \cite{Verduijn2014, Hile2015} or nanoparticles \cite{Frake2015},
the devices which incorporate dopants or nanoparticles still need to be fabricated.
Along with the topographic imaging capability, 
$e$-EFM can be a powerful tool for exploring new nanoscale material systems
such as single molecules, nanoparticle complexes as possible quantum bits.

\section{Conclusion}
We demonstrated that charge sensing by electrostatic force microscopy 
is capable of quantitative electronic energy level spectroscopy 
of quantum dots in three distinct systems, 
epitaxially grown InAs self-assembled QDs on InP, 
colloidal Au nanoparticles on alkanedithiol self-assembled monolayer
and Au nanoislands deposited on NaCl. 

Three different methods for probing the energy level structure of QDs
are demonstrated.
The direction of the temperature-dependent shift of the charging peaks 
in the small tip oscillation case (section~\ref{section:weak_coupling})
and 
the direction of the skewness of the peaks 
in the large tip oscillation case (section~\ref{section:strong_coupling})
enable straightforward identification of shell-filling of QDs.
The energy-dependent tunneling rate which can easily be obtained by $e$-EFM 
offers more general approach to probe energy level structure of the QDs.
Excited-states spectroscopy is also possible 
by measuring the tip oscillation amplitude dependence 
of the single-electron charging spectra.
All the features described above makes $e$-EFM a powerful technique
for readout of the charge state of QDs.

The spatial mapping capability of the $e$-EFM technique provides 
a new route for the research on unconventional QDs 
such as multi-QDs based on nanoparticles,
individual molecules, dopants and defects.
In combination with topographic imaging 
and structural characterization capabilities of AFM, 
$e$-EFM will play an important role in searching for new device elements 
for nanoelectronics including qubits.

\section*{Acknowledgment}
We thank P.~Poole, S.~Studenikin, and A.~Sachrajda at the 
National Research Council of Canada for providing the InAs QD sample and
for fruitful discussion,
L.~Cockins and A.~Tekiel for their experimental contributions,
A.~A.~Clerk and S.~D.~Bennette for their theoretical contributions. 
This work was supported by funding from NSERC and FQRNT.

\section*{References}
%\bibliography{~/Documents/EFS_review}
\bibliographystyle{unsrt}

\end{document}